\documentclass{JHEP3}
\usepackage{amsmath}
\usepackage{epsfig}

\newcommand{\tr} {\operatorname{tr}}
\newcommand{\One} {{\bf 1}} 

\newcommand{\np} {\operatorname{np}}
\newcommand{\cub} {\operatorname{cub}}
\newcommand{\Sma} {\operatorname{Small}}
\newcommand{\Lar} {\operatorname{Large}}

\newcommand{\cO} {{\cal O}}

\newcommand{\La} {{\Lambda}}
\newcommand{\ep} {{\epsilon}}

\newcommand{\ket}[1] {\left|#1\right>}

\title{On the classical equivalence of superstring field theories}

\author{Ehud Fuchs, Michael Kroyter\\
Max-Planck-Institut f\"ur Gravitationsphysik\\
Albert-Einstein-Institut\\
14476 Golm, Germany\\
\email{udif@aei.mpg.de, mikroyt@aei.mpg.de}
}

\abstract{
We construct mappings that send solutions of the cubic
and non-polynomial open superstring field theories to each other.
We prove that the action is invariant under the maps and
that gauge orbits are mapped into gauge orbits.
It follows that the perturbative spectrum around solutions
is the same in both theories.
The mappings also preserve the string field reality condition.

We generalize to the cases of a non-BPS D-brane and of
multi-D-brane systems.
We analyze the recently found analytical solutions of the cubic action,
both in the BPS sector and the non-BPS sector and show that they
span a one parameter family of solutions with empty cohomology
and identical action, which suggests that they are gauge
equivalent. We write the gauge transformations relating these solutions
explicitly. This seems to suggest that open superstring field
theory is able to describe a vacuum solution even around a BPS D-brane.
}

\preprint{AEI-2008-030}

\begin{document}

\section{Introduction}

Covariant open superstring field theory was first constructed by
Witten~\cite{Witten:1986qs}, following his construction of
the bosonic theory~\cite{Witten:1986cc}.
The formal structure of the theories is very similar. They both
rely on the algebraic structures obeyed by the star-product,
the BRST operator $Q$ and the integration over the space of string
fields. There are two new features that arise in the case of
the superstring. One is the existence of
several sectors, namely GSO($\pm$) and NS/R, that should be
separately handled. The other is the notion of
picture number~\cite{Friedan:1985ge}.

Picture number is the expression of the existence of zero modes
for the superghosts of the RNS formalism.
Just as the zero modes of the fermionic $bc$ ghost system imply a
two-fold degeneracy in the description of vertex operators,
namely integrated/unintegrated vertex, the zero modes of the
bosonic $\beta\gamma$ system imply the existence of an infinite
number of equivalent vertices, distinguished by a new quantum number,
called picture number. Picture number is most easily dealt with using
fermionization~\cite{Friedan:1985ge}, where the $\beta,\gamma$ system
is expressed in terms of the $\eta,\xi,\phi$ variables,
\begin{equation}
\beta=e^{-\phi}\partial \xi\,,\qquad \gamma=\eta e^\phi\,.
\end{equation}

The operation of integration over the space of string fields is
realized using a CFT expectation value, which is zero for fields
of zero picture number. This implies that one has either to assign
non-zero picture number to the string field or to append it to
the operations of star-product/derivation/integration.

Witten chose to define the (NS+) string field in the natural $(-1)$
picture and to append a picture changing operator to the cubic term
in the action. This picture changing operator had to be inserted at
the interaction (mid-string) point $z=i$ in order to preserve the
associativity of the star product.
Soon thereafter it was noticed that in
calculations of scattering amplitudes and of iterated gauge
transformations picture changing operators formally
collide~\cite{Wendt:1987zh,Arefeva:1988nn}.
This is an immediate consequence of inserting the operator
at $z=i$, which is invariant under the star-product.
The collisions of picture changing operators produce
singularities, since the OPE of the picture changing operator
with itself is divergent. It was shown that the theory can be
regularized to leading order by adding counter terms (already
at the classical level), but it was also shown that the problems
persist at higher order and it was not understood if and how
could one resolve these problems systematically.

Resolutions of these problems can be achieved by leaving the
string fields at ghost number one, while
modifying their picture number to
zero~\cite{Preitschopf:1989fc,Arefeva:1989cp,Arefeva:1989cm}.
The modified cubic action (in the NS+ sector) reads,
\begin{equation}
\label{CubicAction}
S_{\cub}=- \frac{1}{g_o^2}\int Y_{-2}\Big(\frac{1}{2}\Psi Q\Psi+
            \frac{1}{3}\Psi^3\Big),
\end{equation}
where $g_o$ is the open string coupling constant, $Y_{-2}$
is the double-step inverse picture changing operator
and the string fields are multiplied in the action using Witten's
star product, which we leave implicit throughout the paper.

There are two possible definitions for $Y_{-2}$,
namely the chiral the non-chiral operators.
The chiral is a local operator, whose exact form will not be of
interest for us, while the non-chiral is a non-local operator,
defined using the doubling trick as
\begin{equation}
Y_{-2}(z)=Y(z)Y(\bar z)\,.
\end{equation}
This definition cannot be used on the real line,
where picture changing operators and vertex operators are
usually inserted, due to the singular OPE of $Y$ with itself.
Nonetheless, it makes perfect sense when inserted
in the string mid-point, $z=i$.
In this formalism there are no problems of collisions, since
while the interaction vertex comes with a $Y_{-2}$ factor,
the propagator comes with a factor of its inverse, which
is well defined. These operators are canceled in pairs and
scattering amplitudes can be calculated and give
(at least on-shell and for tree diagrams) the expected
results~\cite{Arefeva:2001ps}.

The two different choices define two a-priori inequivalent
theories. The theory that uses the chiral operator was
shown to have some strange features in level
truncation~\cite{Urosevic:1990as}. Also, being chiral it
does not respect the twist symmetry that one expects it to
obey. For these reasons the non-chiral theory was used in
most studies of cubic superstring field theory and is considered
the more reliable one.
This is the version of the theory we shall be using here.
The non-chiral $Y_{-2}$ was recently used also in the construction
of boundary superstring field theory~\cite{Ishida:2008tj}.

From the action~(\ref{CubicAction}) one derives the equation
of motion
\begin{equation}
\label{YEOM}
Y_{-2}(i)\big(Q\Psi+\Psi^2)=0\,.
\end{equation}
The operator $Y_{-2}(i)$ has a non-trivial kernel. However, this
kernel consists of states of a somewhat singular nature
that are localized at $z=\pm i$.
It is not clear if it has a non-trivial intersection with the
correct space of open string fields, especially since it is
not clear what should this space be. Assuming that this kernel
poses no problem, the equation of motion gets the desired form
\begin{equation}
\label{PsiEOM}
Q\Psi+\Psi^2=0\,.
\end{equation}
The action is invariant under the gauge transformation
\begin{equation}
\label{CubicGauge}
\Psi \rightarrow e^{-\La}(\Psi+Q)e^\La\,.
\end{equation}

Another resolution of the problems, not relying on picture
changing operators, was obtained by
Berkovits~\cite{Berkovits:1995ab}, who developed another form
of superstring field theory. This theory is non-polynomial,
resembling a WZW-theory,
although (unlike closed/heterotic string field
theory~\cite{Zwiebach:1992ie,Okawa:2004ii,Berkovits:2004xh})
string fields are still multiplied using only Witten's
star-product. The non-polynomial theory uses ghost number zero,
picture number zero string fields
that reside in the large Hilbert space. The large Hilbert
space consists of two copies of the small (standard) Hilbert
space, the usual one and another one, which is multiplied by
$\xi_0$. This redundancy is taken care of by a novel gauge
symmetry, which implies that only the copy proportional to
$\xi_0$ is physical. The non-polynomial action reads,
\begin{equation}
\label{BerkoAction}
S_{\np}=\frac{1}{2g_o^2}\oint\Big(
e^{-\Phi} Q (e^\Phi) e^{-\Phi} \eta_0 (e^\Phi)-\int_0^1 dt\,
   e^{-t\Phi} \partial_t e^{t \Phi}
   [e^{-t\Phi}\eta_0 e^{t \Phi},e^{-t\Phi} Q e^{t \Phi}]
\Big),
\end{equation}
and the gauge symmetry is
\begin{equation}
\label{BerGaugeTrans}
e^\Phi \rightarrow e^{-Q\Lambda_Q}e^\Phi e^{\eta_0 \La_\eta}\,.
\end{equation}
In~(\ref{BerkoAction}) and elsewhere we use the square
brackets to denote the graded commutator, meaning that $[A,B]$ is an
anti-commutator if $A$ and $B$ are odd (as it is here)
and a commutator otherwise.
The integral over string fields
corresponds to a CFT expectation value. For the non-polynomial
theory this should be calculated in the large Hilbert space.
To distinguish this case we define
\begin{equation}
\oint A \equiv \left < A \right >_{\Lar}\,,\qquad
\int A \equiv \left < A \right >_{\Sma}\,.
\end{equation}
The integrals are related in a simple way
\begin{equation}
\label{LargeToSmall}
\oint \xi_0 A = \int A\,.
\end{equation}

From the action~(\ref{BerkoAction}) the following
equation of motion is derived,
\begin{equation}
\label{EOMBerk}
\eta_0(G^{-1}Q G)=0\,,
\end{equation}
where we define
\begin{equation}
G=e^\Phi\,.
\end{equation}
Equation~(\ref{EOMBerk}) states that $G^{-1}Q G$ lives in the small
Hilbert space, where the string fields of the cubic theory reside.
It also has the correct quantum numbers to be identified with $\Psi$.
This expression takes exactly the form of a
gauge solution of the cubic theory~(\ref{CubicGauge}).
It is not a genuine gauge
solution, since the ``gauge field'' $G$ lives in the large
Hilbert space. It is nevertheless a solution of the cubic theory.
Thus, there exists a simple map of solutions of the non-polynomial
theory to solutions of the
cubic theory~\cite{Berkovits:2001im,Kling:2002vi},
\begin{equation}
\label{PsiOfG}
\Psi=G^{-1}Q G\,.
\end{equation}

The purpose of this paper is to show that
(at least classically and in the NS sector) the
cubic superstring field theory and Berkovits' non-polynomial
superstring field theory are equivalent.
We prove this in section~\ref{sec:Map}
by adding to the map~(\ref{PsiOfG}) also a map that works in
the other direction. This map also sends solutions to solutions
and leaves the action invariant.
We further show that both maps respect the gauge symmetries,
so we obtain a one to one mapping of gauge orbits of solutions
in both directions.
In section~\ref{sec:Gauge} we show how to write the solutions of
the cubic theory as formal gauge solutions. In this form our map
takes the form of a mapping of gauge fields. We also comment on
bosonic string field theory.
In section~\ref{sec:NonBPS} we describe how the
GSO($-$) sector is added to the theory and to our mapping.
This enables us to generalize the mapping to non-BPS D-branes and
to multi D-brane systems.

In section~\ref{sec:Vacuum} we describe analytical
solutions in both theories on
BPS and non-BPS D-branes that seem to describe the
vacuum without the original D-brane.
These solutions are based on the recent
solutions of~\cite{Fuchs:2007gw,Erler:2007xt,Aref'eva:2008ad}.
While such solutions are to be expected for the non-BPS D-brane,
where a tachyon is present, it may seem strange to have them on
the BPS D-brane, where a physical tachyon is absent. Moreover,
a BPS D-brane carries a RR charge, which is absent in the vacuum
without the D-brane.
Thus, such solutions lead to a change in the closed string
background. We claim, that it seems to be the case
that such solutions are nevertheless realized. We comment on
this issue and offer some other conclusive remarks
in section~\ref{sec:Conclusions}.

\section{Mapping the cubic theory to the non-polynomial one}
\label{sec:Map}

In this section we construct a map of string fields from the cubic
superstring field theory to string fields in the non-polynomial
theory. We construct the map in~\ref{sec:TheMap} and show that
it sends solutions to solution in~\ref{sec:EOM}.
We compose this map and the map~(\ref{PsiOfG}), which goes in
the opposite direction and is defined only for solutions. The
composition in one direction gives the identity transformation on
solutions of the cubic theory and in the other direction it gives
a gauge transformation on the space of solutions of the non-polynomial
theory. Further, we prove in~\ref{sec:GaugeEquiv} that both
maps send gauge equivalent
solutions to gauge equivalent solutions and that all gauge
orbits are accessible.
Thus, the two maps define a one to one correspondence of
gauge orbits of solutions.
A direct consequence of this fact is that the cohomologies
around solutions in both theories agree.
Finally, in~\ref{sec:Action} we study the action of solutions and prove
that it has the same value in both theories.

\subsection{The map}
\label{sec:TheMap}

A map of $\Psi$ to $\Phi$ should contain multiplication by a
$\xi_0=\xi(0)$ factor, which reduces the ghost number by one, while
increasing the picture number by one. In order to get the correct
quantum numbers, we need also an operator that decreases the
picture number by one, without modifying the other quantum numbers,
such as ghost number and conformal weight. Such an operator exists,
namely, the inverse picture changing operator,
\begin{equation}
Y(z)=(c \partial \xi e^{-2\phi})(z)\,.
\end{equation}
Combining the two
operators we get the operator
\begin{equation}
P(z)=(\xi Y)(z)=-(c \xi \partial \xi e^{-2\phi})(z)\,,
\end{equation}
that can be used to define the desired mapping.
So far we did not fix $z$. In fact, we can define a map
\begin{equation}
\label{TheMap}
\Phi=P \Psi\,,
\end{equation}
using a linear combination
\begin{equation}
P = \sum_i k_i P(z_i)\,.
\end{equation}
The operator $P$
contains
$\xi_0$ as long as $\sum k_i\ne0$.
It is also possible to define $P$ using different insertion points
for $Y(z)$ and $\xi(z)$.
This will become useful in~\ref{sec:Action}.

An important property of $P$ is that it is nilpotent,
\begin{equation}
\label{PP0}
P^2=0\,.
\end{equation}
This follows from the OPE
\begin{equation}
P(z)P(w)=-\frac{z-w}{12}
    (c c' \xi \xi' \xi'' \xi''' e^{-4\phi})(w)+\cO\big((z-w)^2\big)\,.
\end{equation}
This is in contrast to the OPE of the inverse picture changing
operator $Y$ with itself,
\begin{equation}
Y(z)Y(w)=-\frac{(c c' \xi ' \xi '' e^{-4\phi})(w)}{(z-w)^2}+
  \cO(\frac{1}{z-w})\,.
\end{equation}
The picture changing operator
\begin{equation}
X(z)=[Q,\xi(z)]
\end{equation}
suffers from a similar singularity. The appearance of these singularities
in Witten's theory were the source of a need for its modification.

We demand that the $P$ operator be nilpotent also under the star-product
\begin{equation}
\label{Nilpotent}
(P\Psi_1) (P\Psi_2) = 0\,.
\end{equation}
We achieve this by inserting $P(z)$ at the string mid-point
\begin{equation}
\label{PlinComb}
P=k_1 P(i)+k_2 P(-i)\,,
\end{equation}
which is invariant under the star-product.

The operator $P(z)$ is the inverse of the BRST charge $Q$ in
the sense that
\begin{equation}
\label{QP1}
[Q,P(z)]=1\,,\qquad \forall z\,.
\end{equation}
We can define a string field
\begin{equation}
A=P(z)\ket{1},
\end{equation}
where $\ket{1}$ is the identity string field. This state
obeys,
\begin{equation}
Q A=\ket{1}.
\end{equation}
From the above it follows that the cohomology of $Q$
is trivial~\cite{Ellwood:2001ig,Ellwood:2006ba} (in the large
Hilbert space). The same also holds for $\eta_0$ that obeys
\begin{equation}
\label{etaXi1}
[\eta_0,\xi(z)]=1\,,\qquad \forall z\,.
\end{equation}

The property~(\ref{QP1}) is crucial for our construction. Thus,
we shall demand that the coefficients in~(\ref{PlinComb})
sum to unity,
\begin{equation}
\label{PkMinusk}
P(k)=k P(i)+(1-k) P(-i)\,.
\end{equation}
Similarly to $P(k)$ we define
\begin{equation}
Y(k)=k Y(i)+(1-k) Y(-i)\,.
\end{equation}

\subsection{Equations of motion}
\label{sec:EOM}

Suppose that $\Phi$ is given by~(\ref{TheMap}), where $P$ is given by
(\ref{PkMinusk}) for some $k$.
The nilpotency of $P$ under the star-product~(\ref{Nilpotent}) implies,
\begin{equation}
\label{GwithP}
G=1+P \Psi\,,\qquad G^{-1}=1-P\Psi\,.
\end{equation}

Let us map $G$ back to $\Psi'$ using~(\ref{PsiOfG}).
For $G$ of the form~(\ref{GwithP}), we find,
\begin{equation}
\label{PsiofG2}
\Psi'=G^{-1}Q G=(1-P\Psi)(\Psi-P Q \Psi)=(1-P\Psi)(\Psi+P \Psi \Psi)=
    \Psi\,.
\end{equation}
Here, we used~(\ref{QP1}), then we used the equation of motion for
$\Psi$~(\ref{PsiEOM})
and the nilpotency of $P$ under the star-product~(\ref{Nilpotent}).
We conclude that the transformation~(\ref{PsiOfG}) is the inverse
of~(\ref{TheMap}) in this direction. This is not true when the
transformations are composed in the other order.

Proving that $G$ defined by our map is a solution is now
straightforward, as
\begin{equation}
\label{EOMtrivial}
\eta_0(G^{-1}Q G)=\eta_0 \Psi=0\,,
\end{equation}
since $\Psi$ resides in the small Hilbert space and
so is annihilated by $\eta_0$.
We conclude that given our map, the equation of motion for $\Phi$
follows from that of $\Psi$.

\subsection{Gauge equivalence of solutions and cohomology}
\label{sec:GaugeEquiv}

Not all solutions of the non-polynomial theory can be written in the
form~(\ref{TheMap}). Here, we prove that all of them are
gauge equivalent to ones that can be so written.
Then, we prove that the two maps~(\ref{TheMap}) and~(\ref{PsiOfG})
map gauge equivalent solutions to gauge equivalent solutions.
Thus, the maps define a one to one and onto mapping of gauge
orbits of solutions of both theories.

Parametrize an arbitrary solution $G$ as
\begin{equation}
G=\frac{1}{1-\Phi}\,.
\end{equation}
In this scheme~(\ref{PsiOfG}) takes the form,
\begin{equation}
\Psi=G^{-1}QG=-QG^{-1}G=Q\Phi \frac{1}{1-\Phi}\,.
\end{equation}
Map it now back to the non-polynomial theory using~(\ref{TheMap}),
\begin{equation}
\label{YAL}
\tilde G=1+PQ\Phi\frac{1}{1-\Phi}=1+\big((1-QP)\Phi\big)\frac{1}{1-\Phi}=
\big(1-Q(P\Phi)\big)G\,.
\end{equation}
We conclude that the state to which $\Psi$ is mapped is gauge equivalent to
the original $G$. Important immediate corollaries are the existence of
a solution of the form~(\ref{TheMap}) in all gauge orbits, the
gauge equivalence of any two solutions that produce the same $\Psi$
via~(\ref{PsiOfG}) and the fact that the various choices of $k$
in~(\ref{PkMinusk}) result in gauge equivalent solutions.

It may seem strange that the gauge transformation relating two
solutions, which give rise to the same solution in the cubic theory, is
the one based on $Q$, present also in the cubic theory.
One should remember, though, that in this equation, $Q$ acts on string
fields residing in the large Hilbert space, which includes more
degrees of freedom. This extra amount of degrees of freedom is fixed
by the other, $\eta_0$-based gauge transformation,
which we did not fix.

Let there be two gauge equivalent solutions of the
non-polynomial theory,
\begin{equation}
\tilde G = e^{-Q \Lambda_Q} G e^{\eta_0 \Lambda_\eta}\,.
\end{equation}
It follows that
\begin{equation}
\begin{aligned}
\tilde G^{-1}Q\tilde G & = e^{-\eta_0 \Lambda_\eta}G^{-1} e^{Q \Lambda_Q}
   Q\big(e^{-Q \Lambda_Q} G e^{\eta_0 \Lambda_\eta}\big)\\ &=
 e^{-\eta_0 \Lambda_\eta}G^{-1}
     \big(Q G e^{\eta_0 \Lambda_\eta}+G Q e^{\eta_0 \Lambda_\eta}\big)=
 e^{-\Lambda}\big(\Psi+Q)e^\Lambda\,,
\end{aligned}
\end{equation}
where we defined
\begin{equation}
\Lambda=\eta_0\Lambda_\eta\,.
\end{equation}
We see that gauge equivalent states are mapped under~(\ref{PsiOfG})
to gauge equivalent states.

Now, let there be two gauge equivalent string fields of the
cubic theory,
\begin{equation}
\tilde \Psi=e^{-\Lambda}\big(\Psi+Q)e^\Lambda\,.
\end{equation}
We get,
\begin{align}
\nonumber
1+P\tilde\Psi &= 1+P\big( -Q e^{-\Lambda}+e^{-\Lambda}\Psi\big)e^\Lambda=
  \big(Q (P e^{-\Lambda}) + P e^{-\Lambda}\Psi\big)e^\Lambda=
  Q (P e^{-\Lambda})\big(1+P\Psi\big)e^\Lambda\\ &
\label{YAL2}
  =Q (P e^{-Q P\Lambda})\big(1+P\Psi\big)e^\Lambda=
  e^{-Q \La_Q}\big(1+P\Psi\big)e^{\eta_0\La_\eta}\,,
\end{align}
where we used~(\ref{QP1}), (\ref{etaXi1}) and defined
\begin{equation}
\label{LaEtaLaQ}
\La_\eta=\xi \La\,,\qquad \La_Q=P\La\,,
\end{equation}
and the location of the $\xi$ insertion is arbitrary.
The result~(\ref{YAL2}) takes the form of a
gauge transformation~(\ref{BerGaugeTrans}) of $(1+P\Psi)$.
We conclude that in this direction gauge equivalent states are
mapped to gauge equivalent states even off-shell.

Let us now consider linearizing the theories around two
equivalent solutions.
The linearized equations of motion of both theories coincide under the maps, 
since the exact equations coincide and similarly for the linearized
gauge transformations. Solutions of the
linearized equation of motion modulo linearized gauge transformations
define the cohomology around the solution. We conclude that the
linearized map around a solution can be interpreted as a map between
the cohomology spaces of the solutions, which is one to one and onto.
This suggests that solutions are mapped to solutions with the same
physical content.

\subsection{The action}
\label{sec:Action}

Another physical property of a solution is its action.
The action of the non-polynomial theory~(\ref{BerkoAction})
can be put in the form, 
\begin{equation}
S_{\np}=\frac{1}{2g_o^2}\oint\Big(
-Q e^\Phi \eta_0 e^{-\Phi} +\int_0^1 dt\,
   \Phi [\eta_0 e^{-t\Phi} , Q e^{t \Phi}] \Big).
\end{equation}
Assume that $\Phi$ is given by~(\ref{TheMap}). We want to prove
that the non-polynomial action coincides with the cubic action
of $\Psi$.

We begin by explicitly writing down,
\begin{equation}
e^{t \Phi}=1+t P \Psi\,,\qquad e^{-t \Phi}=1-t P\Psi\,.
\end{equation}
We use the map to write the first term as
\begin{equation}
S_1\equiv -\frac{1}{2g_o^2}\oint Q G \eta_0 G^{-1}=\frac{1}{2g_o^2}\oint
  (\Psi-P Q \Psi) Y \Psi=-\frac{1}{2g_o^2}\int
  (Y Q \Psi) (Y \Psi)\,.
\end{equation}
Here we used the fact that in the resulting expression $\xi_0$
appears only in $P$, the relation between $P$ and $Y$ and
the relation between the
integration in the large and small Hilbert spaces~(\ref{LargeToSmall}).

We now face a problem. In the above equation we have a collision of
$Y$'s, which is singular.
Let us assume that some sort of a regularization was employed
such that all the expressions are well defined and proceed with
the evaluation of the cubic term (we comment below on the nature of
the needed regularization),
\begin{equation}
\label{S2}
S_2\equiv \frac{1}{2g_o^2}\oint \! \int_0^1 dt\,
   P\Psi [-t Y \Psi , t \Psi-t P Q \Psi]=
   -\frac{1}{6g_o^2}\oint P \Psi [Y\Psi ,\Psi]=
   -\frac{1}{3g_o^2}\int (Y \Psi) (Y \Psi) \Psi\,.
\end{equation}
Here we used~(\ref{Nilpotent}) and evaluated the simple
$t$ integral. Then, we moved to the small Hilbert space and
expanded the commutator giving two identical terms.
Summing up the two terms gives,
\begin{equation}
S_{\np}(k)=S_1+S_2=\int Y(k)^2 L\,,\qquad
  L\equiv -\frac{1}{2g_o^2}
     \Big(\frac{1}{2}\Psi Q \Psi + \frac{1}{3}\Psi^3\Big).
\end{equation}
The action $S_{\np}(k)$ is formally dependent on the choice
of $k$ in~(\ref{PkMinusk}).
However, as we showed in the previous subsection, the choice of $k$
corresponds to a gauge transformation. Hence, the action is actually
$k$ independent.
Notice that the cubic action~(\ref{CubicAction}) is given by
\begin{equation}
S_{\cub}=\int Y_{-2} L\,.
\end{equation}

Consider now an arbitrary $k\neq 0,1$. Evaluating the action one gets,
\begin{equation}
S_{\np}(k)=k^2 S_{\np}(1)+(1-k)^2 S_{\np}(0)+2k(1-k) S_{\cub}\,.
\end{equation}
Using the $k$-independence of the action the above implies,
\begin{equation}
S_{\np}=S_{\cub}\,,
\end{equation}
which ends our proof.

One can get directly $S_{\cub}$ from the non-polynomial theory (plus two
terms that cancel each other) by choosing
\begin{equation}
\label{Special_k}
k=\frac{e^{\frac{i \pi}{4}}}{\sqrt{2}}\,.
\end{equation}
An additional property this $k$ has is that it obeys
\begin{equation}
1-k=\bar k \qquad \Longleftrightarrow \qquad \Re(k)=\frac{1}{2}\,,
\end{equation}
which ensures that the map respects the
string field reality condition.

Before closing this section, let us comment on the
needed regularization.
The points $z=\pm i$ are singular from several points of view:
they are located on the boundary of the local coordinate patch and
they are also invariant under the star product.
It is clear that a regularization would have to move $z$ slightly
away from the local coordinate patch.
One can think of such a regularization as a
regularization of the star product itself,
in the spirit of Witten's original construction.
This interpretation was used in various contexts in recent
developments~\cite{Fuchs:2006gs,Fuchs:2007gw,Kiermaier:2007jg},
where it was represented using
\begin{equation}
\cO\rightarrow s^{-L_0} \cO s^{L_0}\,.
\end{equation}

A proper regularization should always keep some relations
intact, such as those which originate from symmetry principles.
In the case at hand there are two relations that should continue to
hold. First, the nilpotency of $P$ should hold, at least in the limit
in which the regularization is removed, even upon a contraction with
two regularized $Y$ insertions. A direct calculation reveals that
a symmetric limit of colliding two $P$ insertions even with a single
$Y$ insertion is divergent. One way to remedy this problem would be
to use a regularization in which the $Y$ and $\xi$ ingredients of
$P$ approach $\pm i$ at a different pace. It is possible then to
arrange the needed nilpotency in the limit.
Of course, such a regularization can no longer be considered as
related to the star product and we interpret it as defining
a family of regularized solutions.

The other property that the regularization should respect is the
gauge equivalence of regularized
solutions with different $k$ values as well as different $\ep$
values, where $\ep$ is the regularization parameter.
It is clear that for $z\neq \pm i$, more terms should be added
to $G,\,G^{-1}$ as compared with~(\ref{GwithP}).
We believe that such a regularization exists.
We leave finding its explicit form to future work.

\section{Formal gauge form of solutions in the cubic theories}
\label{sec:Gauge}

In this section we discuss the formal gauge representation of
solutions. In~\ref{sec:GaugeForm} we give the formal gauge form
of solutions of the cubic superstring field theory and describe
our map as a map of gauge fields. Then, in~\ref{sec:bosonic}
we discuss the problems in obtaining the same formal representation
for solutions of the bosonic theory and give some directions
towards this end.

\subsection{Gauge form of the solutions}
\label{sec:GaugeForm}

In~\cite{Fuchs:2007gw} a mapping of bosonic solutions, given in a
formal gauge form,
to the non-polynomial theory, was considered.
That mapping could also be used to map superstring fields from the
cubic theory, as we are doing here. 
Let $\Lambda$ be a formal gauge field defining a solution to a cubic
(bosonic or supersymmetric) string field theory via
\begin{equation}
\Psi=e^{-\Lambda}Q e^{\Lambda}\,.
\end{equation}
Define a corresponding solution in the non-polynomial
theory by defining the gauge fields,
\begin{equation}
\label{GaugeMap1}
\Lambda_{\eta} =\xi(z) \Lambda\,,\qquad \Lambda_Q= P(z)\Lambda\,,
\end{equation}
for some $z$. A pure gauge solution of the non-polynomial theory
is given by
\begin{equation}
\label{GaugeMap2}
e^\Phi=e^{-Q\Lambda_Q}e^{\eta_0 \Lambda_{\eta}}=
  e^{P(z)Q\Lambda-\Lambda}e^{\Lambda}\,,
\end{equation}
and we see that the choice of $z$ for $\xi$ drops out of the expression
for the solution, but remains for $P$.
Expansion in powers of $\Lambda$ gives a simple leading order
behaviour,
\begin{equation}
\Phi_1=P(z)Q\Lambda=P(z)\Psi_1\,,
\end{equation}
which has a canonical form (up to the choice of $z$) and agrees
with the first order of~(\ref{TheMap}) (with the same choice of $z$).

We can change our scheme~\cite{Fuchs:2007yy,Fuchs:2007gw} and
redefine
\begin{equation}
1+\Phi=(1-Q\Lambda_Q)\frac{1}{1-\eta_0 \Lambda_{\eta}}=
  (1+P(z)Q\Lambda-\Lambda)\frac{1}{1-\Lambda}=
    1+(P(z)Q\Lambda)\frac{1}{1-\Lambda}\,.
\end{equation}
If we choose to work with a $P$ of the form~(\ref{PkMinusk}), we
can write that as
\begin{equation}
\label{gaugeToSol}
1+\Phi=1+P(Q\Lambda\frac{1}{1-\Lambda})=1+P\Psi\,,
\end{equation}
which is exactly the map~(\ref{TheMap}).

An important question raised in~\cite{Fuchs:2007yy,Fuchs:2007gw}
is how general is the representation of solutions as formal
gauge solutions. We can now give a simple answer for the case of
the cubic superstring field theory: all its solutions can
be formally represented as gauge solutions via~(\ref{PsiOfG}),
where $G$ is defined by~(\ref{GwithP}).

One may wonder what happens when this gauge representation is
used in order to obtain a general
formal gauge form of solutions of the non-polynomial theory
via~(\ref{GaugeMap1}).
Now, the gauge field of the cubic theory is given by
\begin{equation}
\La=\log G=P\Psi\,.
\end{equation}
Plugging this $\La$ into~(\ref{GaugeMap1}) results in an expression,
which generally differs from what one would like to obtain and in
particular is trivial,
\begin{equation}
\label{G1}
\tilde G=1\,,
\end{equation}
if the same $z$ is used in the definition of $P$ and
in~(\ref{GaugeMap1}). This implies that we cannot use our
construction to obtain a general formal gauge representation for
the non-polynomial theory. This should have been expected, since
we did not extend the space in any way that would allow the
introduction of formal gauge fields.

\subsection{Formal gauge representation for the bosonic solutions}
\label{sec:bosonic}

The map of gauge fields to the non-polynomial
theory~(\ref{GaugeMap1}), (\ref{GaugeMap2}) can also be used to map
bosonic string fields to the non-polynomial theory\footnote{The idea of
mapping bosonic solutions to supersymmetric ones may seem strange,
since the two theories are obviously different. However, it seems
natural to assume that a bosonic string field, without an explicit
dependence on more than ten space-time directions, can be mapped to
some sort of a supersymmetric string field, whose physical meaning
is similar. Using such a map we would be able to learn about the
superstring from its simpler bosonic counterpart. For another
relation of the two theories see~\cite{Berkovits:1993xq}.}.
On the other hand, the mapping~(\ref{TheMap}) cannot
be generally used to map bosonic solutions to analogous ones in
the non-polynomial theory. The reason being that the supersymmetric
BRST charge contains terms that do not exist in the bosonic theory,
\begin{equation}
Q=Q_0+Q_1+Q_2\,.
\end{equation}
Here, the subscript of $Q$ counts its $\phi$ charge. The bosonic
BRST charge is contained in $Q_0$.
The difference $Q_0-Q_{bos}$ is zero when acting upon bosonic states
not containing factors of the $b$ ghost. The $Q_2$ term, on the other hand,
gives zero upon terms not containing the $c$ ghost.
The $Q_1$ term is zero on terms not containing the $\partial X^\mu$
oscillators. Thus, bosonic
solutions are generally modified when translated to the analogous
supersymmetric ones.
This implies that~(\ref{PsiofG2}) does not simply generalize to the
bosonic case, since the use there of the supersymmetric $Q$ is
important for its derivation. We do not have a general from for a
formal gauge representation for bosonic solutions.

One can still map bosonic solutions to supersymmetric
ones, in those cases where a formal gauge representation
of the solution exists,
\begin{equation}
\Psi_{bos}=e^{-\La_{bos}}Q_{bos} e^{\La_{bos}}\,.
\end{equation}
One can seek a formal gauge representation for bosonic solutions
by enlarging the bosonic Hilbert space using a
couple of canonically conjugated anti-commuting variables
$\tilde b_0,\tilde c_0$. Define
their conformal weight to be zero and their ghost number to be
$-1$ and $1$ respectively. Also, define
\begin{equation}
\tilde c_0 \ket{0}=0\,.
\end{equation}
Then, the redefinition
\begin{equation}
\tilde Q_{bos} = Q_{bos}+\tilde c_0\,,
\end{equation}
does not change the cohomology problem when restricted to the subspace
without the $\tilde b_0$ mode. In the enlarged space, on the other hand,
$\tilde Q_{bos}$ has a trivial cohomology,
\begin{equation}
\tilde Q_{bos} \Psi=0 \quad \Rightarrow \quad \Psi=Q (\tilde b_0 \Psi)\,.
\end{equation}
We also define that these operators simply multiply with respect to the star
product. Then, we can write all bosonic solutions as formal gauge solutions,
\begin{equation}
\Psi=G^{-1} \tilde Q_{bos} G\,,\qquad G=1+\tilde b_0 \Psi\,.
\end{equation}

While the above may be useful for the bosonic theory, it does not help
us with mapping bosonic solutions to supersymmetric ones. A
redefinition of the supersymmetric $Q$ according to
\begin{equation}
\tilde Q=Q+\tilde c_0\,,
\end{equation}
may be used to define a supersymmetric string field $G^{-1} \tilde Q G$.
Unfortunately, this string field contains in general a $\tilde b_0$ piece,
due to the $Q-Q_{bos}$ term. So it cannot be used for deriving
supersymmetric solutions from the bosonic ones.

\section{Incorporating the GSO($-$) sector}
\label{sec:NonBPS}

Our construction can be generalized to the cases of non-BPS
D-branes and D-brane systems. Recall first that in order to describe the
GSO($-$) sector one has to introduce into the theory the so called
``internal Chan-Paton indices''~\cite{Berkovits:2000hf,Arefeva:2002mb}.
For the non-polynomial theory one has to tensor the GSO(+) sector
with the two dimensional unit matrix $\One$ and the GSO($-$) sector with
the Pauli matrix $\sigma_1$. The two gauge fields of a GSO(+) sector
are to be tensored with $\sigma_3$, while the gauge fields of the GSO($-$)
sectors get a factor of $i\sigma_2$. The operators $Q,\eta_0$ are
also tensored with $\sigma_3$ and in the action a trace over the
new matrix space should be performed and the result should be normalized
by $2=\tr(\One)$. With these simple rules all the algebraic properties
hold just as in the case of a single GSO(+) sector. The non-BPS
D-brane can be described by introducing one GSO(+) and one GSO($-$)
sector, while for more complicated brane systems one has to introduce
in addition to the internal space also the genuine Chan-Paton space,
with each factor tensored also in the internal space according to
its GSO parity. The same story works for the cubic theory. Now,
however, the gauge fields are the ones that get the factors of
$\One,\sigma_1$ and the string fields are tensored with
$\sigma_3,i\sigma_2$. The only other modification is the inclusion
of a factor of $\sigma_3$ also for $Y_{-2}$.

The algebraic properties that are important for our construction
are the commutation
relations~(\ref{QP1},\ref{etaXi1}). To ensure that they
continue to hold we have to assign a factor of $\sigma_3$
to $P(z)$ and $\xi(z)$, which implies that $Y(z)$ carries a
factor of $\One$. Then, upon moving from the large to the small Hilbert
space by integrating the $\xi_0$, the leftover $\sigma_3$ factor
combines with the two $Y$'s and the equality of the action also follows.
With these simple assignments all the proofs of section~\ref{sec:Map}
carry over to the general brane system without any further modification.
We conclude that there is no obstruction for generalizing our
map to an arbitrary D-brane system.

\section{Vacuum solutions}
\label{sec:Vacuum}

In this section we discuss vacuum solutions in superstring field theory.
In~\ref{sec:VacuumBPS} we describe the preliminary proposal for a vacuum
solution in the non-polynomial theory on a BPS D-brane
given in~\cite{Fuchs:2007gw}.
We explain the problem of the proposal and correct it. Then, we discuss
the analogous construction, by Erler, in the cubic theory~\cite{Erler:2007xt}
and show that the two are related by~(\ref{TheMap}). This proves that the
solution of the non-polynomial theory shares the properties of correct
action and zero cohomology derived in~\cite{Erler:2007xt} for the cubic
solution.

The solution on the BPS D-brane was criticized in~\cite{Aref'eva:2008ad},
where a solution living on a non-BPS D-brane was constructed.
In~\ref{sec:VacuumNonBPS} we describe the criticism and the new solution
and prove that the cohomology of the solution is trivial as expected.
We further study the one parameter family of solutions interpolating
Erler's solution on the non-BPS D-brane and the new solution and show
that all the members in this family are gauge equivalent.

\subsection{On the BPS D-brane}
\label{sec:VacuumBPS}

In~\cite{Fuchs:2007gw}, it was suggested that the map of gauge
solutions~(\ref{GaugeMap1}),(\ref{GaugeMap2}) from the bosonic
to the non-polynomial theory can be used
to find a ``tachyon vacuum'' in the supersymmetric theory. Since
there is no tachyon on the BPS D-brane the new vacuum cannot
represent tachyon condensation. It can still, potentially,
represent a state where the D-brane is absent. 

During most of that work the operator $P(0)$ was used. This results in
\begin{equation}
\Lambda_Q=(P B c)(0)\ket{0}=P(0)\ket{0},
\end{equation}
implying that
\begin{equation}
Q\Lambda_Q=\ket{0},
\end{equation}
which lives in the small Hilbert space.
The gauge transformation $\eta_0\Lambda_{\eta}$ always leads,
by definition, to a state in the small Hilbert space.
The solution resides in the small Hilbert space and is therefore pure gauge.
Since the source of illness here is the OPE of $P(z)$ and $B c(z)$, an
immediate remedy would be to move $P(z)$ to another point. Here, we suggest
to move it to $\pm i$, that is, to use $P$ of~(\ref{PkMinusk}),
so as to get
\begin{equation}
\label{FKmodified}
\Lambda_Q=P\Lambda\,.
\end{equation}
This expression lives in the large Hilbert space.

In order to calculate the coefficients of the various fields
we have to expand $P$ around
$z=0$. We choose $k$ of~(\ref{Special_k}) and use a derivative,
i.e., conformal weight expansion,
\begin{equation}
\label{Pexpand}
P=(\cos \partial - \sin \partial)P(0)=P(0)-P'(0)-\frac{1}{2}P''(0)+\ldots\,.
\end{equation}
To find the lowest non-trivial field we can use~(\ref{gaugeToSol}) and
recall that for $\Psi$ the lowest level field with non-zero value is
$c(0)\ket{0}$ and the next non-zero fields are of two levels higher.
Acting with~(\ref{Pexpand}) on $c(0)\ket{0}$ we get
\begin{equation}
P c(0)\ket{0}\approx -P'(0) c(0)\ket{0}=-c c' \xi \xi' e^{-2\phi}\ket{0}.
\end{equation}
The field $c c' \xi \xi' e^{-2\phi}$ lives in the large Hilbert space,
so it may give rise to some physical content.
It does not obey the Siegel gauge, so it was not studied much in the
level-truncation literature.
It may seem to describe an unintegrated vertex operator
because of the $c$ factor. However, a direct evaluation reveals that
it does not correspond to a vertex operator. This field is the only field
other than the identity field with all quantum numbers being zero.
It follows that there is no way to correct it in order to obtain a
vertex operator so it is an auxiliary field.

In~\cite{Erler:2007xt}, Erler suggested to use the trivial mapping of
Schnabl's gauge field to the cubic superstring field theory, in order
to obtain a non-perturbative vacuum in this theory. He further showed
that it gives the expected action of a solution describing the absence
of the D-brane and that the cohomology around the solution is empty.
The cohomology proof is in fact straightforward. The resulting string
field is almost identical to the bosonic one and differs only due to
the new terms in $Q$. This gives,
\begin{equation}
\label{ErlerSol}
\Psi=\Psi_{bos}+B \gamma^2\ket{0}=
   \Psi_{bos}+B \eta\partial \eta e^{2\phi}\ket{0}.
\end{equation}

The cohomology proof in the bosonic case~\cite{Ellwood:2006ba} used
the operator
\begin{equation}
\label{A}
A=B \int_0^1 dt\, \Omega^t\,,
\end{equation}
where $\Omega=\ket{0}$ is the vacuum and its powers are
the wedge states $\Omega^t=\ket{t+1}$~\cite{Rastelli:2000iu}.
This field obeys the relation
\begin{equation}
\label{ES}
Q_{\Psi} A=QA+\Psi A+A\Psi=1\,,
\end{equation}
which implies that the cohomology is empty. Since the only new term
in~(\ref{ErlerSol}) is simply proportional to $B$ and contains no $c$
factors, the above
equation holds without modifying $A$ for the supersymmetric case as well.

The calculation of the action relies on the fact
that in order to get a non-trivial result after the integration, the integrand
has to be proportional to $e^{-2\phi}$. The picture changing operator
$Y_{-2}$ supplies a $-4$ factor to the $\phi$ momentum in the action,
so the string field should supply a factor of two. This can come either
from the $\gamma^2$ term in~(\ref{ErlerSol}) or from $Q$ acting on $c$.
Finally, all expectation
values give linear functions rather than the trigonometric functions
one gets in the bosonic case. This fact enables a much simpler evaluation,
which results in the correct D-brane tension.

Mapping Erler's solution using~(\ref{TheMap}) is equivalent to using the
modified gauge representation~(\ref{FKmodified}). From the discussion
of section~\ref{sec:Map} we conclude that the solution of the
non-polynomial theory also has the correct action and cohomology.

\subsection{On the Non-BPS D-brane}
\label{sec:VacuumNonBPS}

In~\cite{Aref'eva:2008ad}, Aref'eva, Gorbachev and Medvedev (hereafter AGM),
criticized Erler's work.
First, it was claimed that a vacuum solution on a non-BPS D-brane should
non exist. Then, it was claimed that the condensation of the string field
$c(0)\ket{0}$, which is the leading order in Erler's solution, should not
correspond to the tachyon vacuum anyway, since this field, called
``the BPS tachyon'' is an auxiliary field and not a vertex operator and
is an artifact of working in the zero picture. It is easy to see that
if one tries to map it to the ``natural'' $-1$ picture, it is mapped to zero.
In fact, a solution based on this vertex operator (in the theory
with chiral $Y_{-2}$) was found using
lowest-level truncation in~\cite{Arefeva:1990ei}, where it was interpreted
as a supersymmetry breaking solution.
In AGM it was suggested that the correct resolution
of these problems is to define the string field on the non-BPS D-brane
using a gauge field with Erler's gauge field in the GSO(+)
sector\footnote{A GSO(+) component is indispensable, since the action of
a purely GSO($-$) solution is trivially zero.} and with
\begin{equation}
\label{GSO-Gauge}
\Lambda_-=B \gamma(0)\ket{0},
\end{equation}
in the GSO($-$) sector.

It is immediate that the action is the same as in Erler's case.
We think of the AGM solution as being expanded around Erler's one
and write the difference
\begin{equation}
\tilde \Psi=\Psi_{AGM}-\Psi_E\,.
\end{equation}
From general considerations it follows that
\begin{equation}
\label{AGMaction}
S_{AGM}=S_E+\frac{1}{6}\int \frac{1}{2}\tr (Y_{-2} \tilde\Psi^3)\,.
\end{equation}
The form of the gauge field~(\ref{GSO-Gauge}) implies that
\begin{equation}
\label{tildePsi+}
\tilde\Psi_+ \approx \gamma B \gamma\,,
\end{equation}
carries momentum two, while the GSO($-$) part given by
(split-string notations~\cite{Okawa:2006vm,Erler:2006hw,Erler:2006ww})
\begin{equation}
\label{tildePsi-}
\tilde\Psi_-=F c \frac{K B}{1-\Omega}\gamma F +
    F\gamma \frac{K B}{1-\Omega}c F\,,
\end{equation}
carries one unit of $\phi$ momentum.
It follows that the minimal $\phi$-momentum power that one can get in
the above action is four, which implies
\begin{equation}
S_{AGM}=S_E\,.
\end{equation}

The proof (not given in AGM) that the cohomology of the AGM
solution is empty relies again on the same $A$~(\ref{A})
as in the previous cases, only now it should be tensored with
$\sigma_3$. For~(\ref{ES}) to hold in this case, we only have to prove
\begin{equation}
[A,\tilde\Psi]=0\,.
\end{equation}
This is trivial for the GSO(+) part~(\ref{tildePsi+}), which
is proportional to $B$, without any $c$ insertions.
The GSO($-$) part~(\ref{tildePsi-}) gives
\begin{equation}
[\tilde\Psi,A]=\Big(-F\gamma \frac{K B}{1-\Omega}c F B \int_0^1 \Omega^t+
     \int_0^1 \Omega^t B F c \frac{K B}{1-\Omega}\gamma F\Big)
  \otimes \sigma_1\,.
\end{equation}
Expanding the denominator we can write the terms in parentheses as
\begin{equation}
\begin{aligned}
\nonumber
K & B \int_0^\infty
   \big(\Omega^t (\gamma(0)\ket{0})-(\gamma(0)\ket{0}) \Omega^t\big)=
B \int_0^\infty \partial_t
   \big(\Omega^t (\gamma(0)\ket{0})-(\gamma(0)\ket{0}) \Omega^t\big)=\\
& B\big( \Omega^\infty (\gamma(0)\ket{0})
      -(\gamma(0)\ket{0}) \Omega^\infty\big)=0\,,
\end{aligned}
\end{equation}
where we used the fact that $K$ acts as a derivation with respect to the
strip length and the final identity is correct, upon a contraction with a
Fock space state, to order $N^{-3}$, where
we used $\Omega^N$ with $N\rightarrow \infty$ for regularization.
This is the same behaviour as in the original proof of the vanishing of the
cohomology~\cite{Ellwood:2006ba}.

The calculation of the action, the triviality of the cohomology
and the existence of $\gamma$
in the gauge field seem to be in favour of the interpretation of
the AGM solution as representing tachyon condensation on the non-BPS
D-brane. Nevertheless, one may wonder whether this solution
is indeed different from Erler's one, or are they gauge equivalent.
Furthermore, the reasoning regarding the form of the solution
as well as the evaluation of the action and the cohomology
remains unchanged if one
multiplies the GSO($-$) gauge field $\Lambda_-$
by an arbitrary parameter $\ep$,
\begin{equation}
\label{AGMmodified}
\Lambda_+ = B c(0)\ket{0},\qquad
\Lambda_- = \ep B \gamma(0)\ket{0}.
\end{equation}
In this manner one gets a one parameter family $\Psi_{\ep}$ of
solutions generated by~(\ref{AGMmodified}). This family interpolates
between Erler's solution and the AGM solution.
All these solutions have the same action and trivial cohomology
so it is natural to expect them to be
gauge equivalent. If this is indeed the case, one is to
conclude that the $\ep=0$ solution can also be used to describe
tachyon condensation on the non-BPS D-brane. It is then very
natural to assume that Erler's solution on the BPS D-brane
and its counterpart in the non-polynomial theory
indeed manage to describe the state without the BPS D-brane despite
the fact that there are no tachyons living on it.

Writing the gauge transformation between two solutions in the
$\ep$-family is (at least formally) immediate. Since these
solutions are formally given as gauge solutions one only has
to compose these two gauge transformation to get a new one.
Then, the question is whether the resulting
transformation is a genuine gauge transformation or whether it is
singular in any sense.

We call the gauge transformation from Erler's solution, with
$\ep=0$ to the $\ep$-solution $\Psi_{\ep}$, $e^{\Lambda_E^{\ep}}$.
Recall however that the gauge form of the solutions is given
in the left scheme, so we can write
\begin{equation}
\begin{aligned}
\label{EepGauge}
e^{\Lambda_E^{\ep}}& =(1-\Lambda_E)\frac{1}{1-\Lambda_{\ep}}=
  1\otimes\One+\ep B \gamma(0)\ket{0}\otimes\sigma_1\,,\\
e^{-\Lambda_E^{\ep}}& =(1-\Lambda_{\ep})\frac{1}{1-\Lambda_E}=
  1\otimes\One-\ep B \gamma(0)\ket{0}\otimes\sigma_1\,,
\end{aligned}
\end{equation}
where we used the relation
\begin{equation}
\label{LambdaLambdaZero}
\Lambda_-^2=\Lambda_- \Lambda_+=0\,,
\end{equation}
to get the final expressions.
These transformations form an abelian group with the simple
multiplication rule,
\begin{equation}
e^{\Lambda_E^{\ep}}e^{\Lambda_E^{\tilde \ep}}=
  e^{\Lambda_E^{\ep+\tilde \ep}}\,,
\end{equation}
where we again used~(\ref{LambdaLambdaZero}).
We see no sign of a singularity at any value of $\ep$.

We can also check that this gauge transformation acts
correctly on $A$
\begin{equation}
A_{\ep}=e^{-\Lambda_E^{\ep}}A \, e^{\Lambda_E^{\ep}}=A\,.
\end{equation}
A singularity in the gauge transformation~(\ref{EepGauge}) would
probably express itself as a singularity in the above equation.
This does not happen. Finally, one may want to calculate some
invariants of the $\Psi_{\ep}$ solution as
in~\cite{Ellwood:2008jh,Kawano:2008ry} and verify that they are
$\ep$-independent. We did not try this method, but we believe that
if tried it would indeed result in $\ep$-independent expressions.

We can map the $\ep$-family to the non-polynomial
theory. Explicitly, the solution
\begin{equation}
\begin{aligned}
\label{epSolution}
\Psi_\ep=\Big(F c \frac{K B}{1-\Omega}c F +
    F\gamma \big(\frac{\ep^2 K}{1-\Omega}+(1-\ep^2)\big) B\gamma F\Big)&
     \otimes\sigma_3\\+\,
 \ep\Big(F c \frac{K B}{1-\Omega}\gamma F +
    F\gamma \frac{K B}{1-\Omega}c F\Big)& \otimes(i\sigma_2)\,,
\end{aligned}
\end{equation}
is mapped to
\begin{equation}
\begin{aligned}
\label{nonPolyTach}
\Phi_\ep=P\Big(F c \frac{K B}{1-\Omega}c F +
    F\gamma \big(\frac{\ep^2 K}{1-\Omega}+(1-\ep^2)\big) B\gamma F\Big)&
     \otimes\One\\+\,
 \ep P\Big(F c \frac{K B}{1-\Omega}\gamma F +
    F\gamma \frac{K B}{1-\Omega}c F\Big)& \otimes\sigma_1\,.
\end{aligned}
\end{equation}
While this mapping seems to be trivial, it serves us to address one last
important point.
The $B$ line integral runs from $-i$ to $i$ and the
$B$ operator has non-trivial commutation relation with $P$.
It follows that again we have to regularize in order to make sense of
the expression. We let $B$ pass to the right of the regularized $P$
insertion, as is implicit in~(\ref{nonPolyTach}). Then, one may wonder
whether the commutation relation of $B$ and $P$ would invalidate the
nilpotency of $\Phi$, which was inherited from that of $P$.
As in~\ref{sec:Action}, we can establish the nilpotency by defining a
different pace for the regularization of $Y(z)$ and $\xi(z)$.
All our treatment remains consistent when such a
regularization is used.

When using wedge state algebra in the evaluation of specific expressions,
one has to include the so called phantom-terms.
The form and origin of these terms is quite clear
by now~\cite{Erler:2007xt}.
They originate from replacing terms of the form $(1-\Omega)^{-1}$
with their series form.
The form and amount of phantom terms needed depends on the specific
solution and on the theory considered.
For~(\ref{epSolution}), it was found that two phantom terms are needed
and their form was explicitly given in~\cite{Erler:2007xt,Aref'eva:2008ad}.
Due to the linear nature of the mapping,
one can then use this result, in order to write down also the
phantom pieces needed for~(\ref{nonPolyTach}).
Of course, the phantom pieces are an artifact of a specific regularization.
There is no need for the phantom piece, neither in~(\ref{epSolution})
nor in~(\ref{nonPolyTach}) when using other consistent regularizations,
such as level-truncation. The advantage of using the wedge-state based
regularization is the ability to obtain analytical results
(in some calculations).

\section{Conclusions}
\label{sec:Conclusions}

In this work we presented mappings between gauge orbits of the cubic and
the non-polynomial superstring field theories.
The mapping in one of the directions depends on a parameter $k$, which
turns out to be a gauge degree of freedom. One can choose $k$
in a way that respects the string reality condition.
An obvious question is whether it is possible to extend our construction
to the quantum theory.
Two important challenges stand in the way: the extension of the maps to
general (off-shell) string fields and the inclusion of the Ramond sector.

The space of string fields that one has to map for the off-shell theory
is larger than the space of string fields we considered in this work.
This stems from the fact that both theories are infinitely reducible
gauge theories. Moreover, the reducibility of the gauge symmetry is
realized only on-shell. This suggests that quantization should be
addressed using the BV
formalism~\cite{ZinnJustin:1974mc,Batalin:1981jr,Batalin:1984jr}
(see~\cite{Henneaux:1992ig,Gomis:1994he} for reviews),
as was done for the bosonic theory
in~\cite{Bochicchio:1986zj,Bochicchio:1986bd,Thorn:1986qj}\footnote{A
BRST study of the bosonic string was performed
in~\cite{Erler:2004hv,Erler:2004sy}.}.
The enlargement of the space of fields comes from the inclusion of
the (second quantized) ghosts and antifields. It can happen that
the maps can be defined in this enlarged space despite not being
defined on the smaller space. These hypothetical maps would have
to commute with the superbrackets and would therefore have the
interpretation of canonical transformations.

The fact that the gauge
symmetries of the two theories are different may be an obstacle for
this ambitious plan. A possible resolution is to gauge fix one of the
gauge symmetries of the non-polynomial theory. One can find in the
literature the gauge fixing
\begin{equation}
\xi_0 \Phi=0\,.
\end{equation}
This is not invariant under the star product.
A more adequate gauge choice seems to be
\begin{equation}
\xi(i) \Phi=0\,,
\end{equation}
or the analogous form in the physical case of a
theory with the reality condition included. This gauge fixing is
immediately solved by writing
\begin{equation}
\Phi=\xi(i) \tilde \Phi\,,
\end{equation}
and our map reduces in this case to
\begin{equation}
\tilde \Phi=Y(i) \Psi\,.
\end{equation}
Now, the spaces and gauge symmetries of both theories are quite
similar.
For the theory with the reality condition imposed, this construction
does not work in this way, since $P$ of~(\ref{PkMinusk})
cannot be written as
a product of $\xi$ and $Y$. One possible resolution of this problem is
to impose the gauge condition directly with $P$, with the immediate
solution being~(\ref{TheMap}). The fact that the transformation
is linear implies that at least formally no problematic measure
factors would emerge. The question is how to generalize the inverse
map~(\ref{PsiOfG}) so as to include also off-shell states and anti-fields.

The inclusion of the Ramond sector is essential,
since Ramond states appear in loops even in scattering processes
of NS states. This raises the familiar problems related
to defining an action for the Ramond sector in superstring field
theory~\cite{Berkovits:2001im,Michishita:2004by}.
We hope that it will be possible to extend this work to the Ramond
sector. Presumably it could also help to resolve some of the problems
related to the definition of the Ramond sector action.

Understanding the Ramond sector is desirable already at the classical
level. For one thing, one may want to be able to study classical
Ramond sector solutions. Another important issue is the
evaluation of the complete spectrum of the theory around a classical
solution. As we mentioned in~\ref{sec:VacuumNonBPS}, a preliminary
numerical study of a solution analogous to Erler's one performed
in~\cite{Arefeva:1990ei} resulted in the interpretation of the
new vacuum as a supersymmetry breaking one. Understanding the
Ramond sector will enable us to study this question analytically.
We believe that this solution represents the closed string
vacuum and so should have a vanishing cohomology also in the Ramond
sector. Not only that the supersymmetry should not be broken around
it, but the solution should also be left invariant under a larger
supersymmetry. Other aspects of supersymmetry are also of importance.

A common objection to the use of the cubic theory is its usage of
the (inverse) picture changing operators $Y(\pm i)$. The problem
being that these operators have non-trivial kernels, which may
result in the equations of motion~(\ref{YEOM}) and (\ref{PsiEOM}) being
inequivalent.
It is not clear if this is really a problem, since the problematic
states are of a somewhat exotic form, i.e., they have a factor that
is localized at the string mid-point and it is not clear if such
states belong to the space of string fields that one has to consider.
We used just these states in our map to the non-polynomial theory.
Hence, we may conclude that the cubic theory on a space not including
these states is equivalent to the non-polynomial one on a space where
they are included.
One may still wonder whether it is possible to define an adequate
space of string fields for the cubic theory. It is known for example
that the space of finite combinations of local insertions on the
vacuum is not closed under the star product. Closeness under the
star product is certainly a very desirable property of the looked-for
space. Can one define a space closed under the star product, that is
large enough for including the expected classical solutions, yet
not containing states with insertions at $\pm i$?

In order to get some intuition about the needed space of string fields
we may use the remarkable developments in the field during the last
couple of years, following Schnabl's analytical
solution~\cite{Schnabl:2005gv}. Subsequent
work~\cite{Okawa:2006vm,Fuchs:2006hw,Ellwood:2006ba} enabled the
proof of Sen's conjectures~\cite{Sen:1999mh,Sen:1999xm}, analytical
solutions describing marginal deformations were
found~\cite{Schnabl:2007az,Kiermaier:2007ba,Erler:2007rh,Okawa:2007ri,Okawa:2007it,Fuchs:2007yy,Fuchs:2007gw,Kishimoto:2007bb,Kiermaier:2007vu,Kiermaier:2007ki},
scattering amplitudes and gauge conditions were
studied~\cite{Fuji:2006me,Rastelli:2007gg,Kiermaier:2007jg,Kiermaier:2008jy}
and much
more~\cite{Fuchs:2006an,Rastelli:2006ap,Okawa:2006sn,Bonora:2006tm,Ellwood:2007xr,Kwon:2008ap,Ishida:2008jc,Kawano:2008jv,Bagchi:2008et,Hellerman:2008wp}.
These developments used to a large extent
wedge states~\cite{Rastelli:2000iu,Rastelli:2001vb} with local
insertions~\cite{Schnabl:2002gg} and with insertions of the line integrals
$B,K$. It seems plausible that the space of string fields should be based
on such states, which do form a closed algebra under the star product
(similar but extended subalgebras may include general
surface states~\cite{LeClair:1989sp,LeClair:1989sj} with local insertions
and the subalgebra of wedge and butterfly
states~\cite{Fuchs:2004xj,Uhlmann:2004mv} with local insertions).
Of course, infinite linear combinations are still needed.
This may result in no restriction at all on the space, since without
restricting the coefficients all possible states can be formed from local
insertions, even without the introduction of wedge states. Thus, we have
to assume that the coefficients should be restricted not to grow, in
some sense, faster than in the known cases. If this is the case, the
formal problems in our construction can be solved, as was explained
at the end of~\ref{sec:VacuumNonBPS}.
Obviously, more research in this direction is needed.

We showed that the vacuum solution on a non-BPS D-brane is gauge equivalent
to a solution that can be defined also on the BPS D-brane.
This convinced us that the solution
for the BPS D-brane should be taken seriously.
The main criticism on this solution is that it contradicts our understanding
that the BPS D-brane is stable. In other words, a condensation of a BPS D-brane
would correspond to a non-continuous change in the Ramond-Ramond charge.
Yet, we can think of a continuous dynamical process in which a D-brane disappear.
For example, we can start with a system of two D5-branes, send one D5-brane to 
spatial infinity and remain with only one D5-brane.
One might worry that hiding one D-brane at infinity is only a trick and
try considering a compact space.
However, on a compact space the total RR change has to vanish.
This means that the possibility of hiding the D-brane at infinity is
correlated with the existence of a non-vanishing RR charge.
We could also consider space-time filling D9-branes.
In this case tadpole cancellation would constrain the total charge to zero.
All the above arguments are related to the close-string sector, which
we did not consider in this paper.
Still, it is important to show that they work out correctly, otherwise
inconsistencies would probably show up at 1-loop calculations.
A related issue is the common expectation that open string field
theory should describe only changes of open string moduli.
It would be very desirable to understand better the realm of open string
field theory.

\section*{Acknowledgments}

We would like to thank the organizers of the workshop
``String field theory and related aspects'' held at the
Arnold Sommerfeld Center, Munich, for hospitality and for
providing such a stimulating research environment. We thank the
participants of this conference for many discussions.
We would also like to thank Ido Adam, Guillaume Bossard,
Stefan Fredenhagen and Stefan Theisen for useful discussions.
The work of M.~K.\ was supported by a Minerva fellowship
during the initiation of this project.
This research was supported by the German-Israeli Project
cooperation (DIP H.52).

\newpage
\bibliography{FK}

\end{document}